\documentclass[twoside,twocolumn]{article}

\usepackage{blindtext} % Package to generate dummy text throughout this template 

\usepackage{graphicx}
\usepackage{subfigure}

\usepackage[sc]{mathpazo} % Use the Palatino font
\usepackage[T1]{fontenc} % Use 8-bit encoding that has 256 glyphs
\usepackage{microtype} % Slightly tweak font spacing for aesthetics

\usepackage[english]{babel} % Language hyphenation and typographical rules

\usepackage[hmarginratio=1:1,top=32mm,columnsep=20pt]{geometry} % Document margins
\usepackage[hang, small,labelfont=bf,up,textfont=it,up]{caption} % Custom captions under/above floats in tables or figures
\usepackage{booktabs} % Horizontal rules in tables

\usepackage{lettrine} % The lettrine is the first enlarged letter at the beginning of the text

\usepackage{enumitem} % Customized lists
\setlist[itemize]{noitemsep} % Make itemize lists more compact

\usepackage{abstract} % Allows abstract customization
\usepackage{titlesec} % Allows customization of titles
\renewcommand\thesection{\Roman{section}} % Roman numerals for the sections
\renewcommand\thesubsection{\roman{subsection}} % roman numerals for subsections
\titleformat{\section}[block]{\large\scshape\centering}{\thesection.}{1em}{} % Change the look of the section titles
\titleformat{\subsection}[block]{\large}{\thesubsection.}{1em}{} % Change the look of the section titles

\usepackage{fancyhdr} % Headers and footers
\usepackage{titling} % Customizing the title section

\usepackage{hyperref} % For hyperlinks in the PDF

\newcommand{\pr}{\partial}

\newcommand{\ep}{\epsilon}

\newcommand{\om}{\omega}

\newcommand{\beq}{\begin{equation}}
\newcommand{\eeq}{\end{equation}}

\newcommand{\ball}{\begin{align}}
\newcommand{\eall}{\end{align}}

\newcommand{\beqar}{\begin{eqnarray}}
\newcommand{\eeqar}{\end{eqnarray}}

\newcommand{\ben}{\begin{enumerate}}
\newcommand{\een}{\end{enumerate}}

\pretitle{\begin{center}\huge\bfseries} % Article title formatting
\posttitle{\end{center}} % Article title closing formatting
\title{The story of magnetism: from Heisenberg, Slater, and Stoner to Van Vleck, and the issues of exchange and correlation} % Article title
\author{%
\textsc{Navinder Singh}\thanks{Cell: +919662680605; Landline: 00917926314457.} \\[1ex] % Your name
\normalsize Physical Research Laboratory, Ahmedabad, India. \\ % Your institution
\normalsize \href{mailto:navinder.phy@gmail.com}{navinder.phy@gmail.com} % Your email address
}
\date{\today} % Leave empty to omit a date
\renewcommand{\maketitlehookd}{%
\begin{abstract}
\noindent This article is devoted to the development of the central ideas in the field of magnetism. The presentation is semi-technical in nature and it roughly follows the chronological order.  The key contributions of Van Vleck, Dorfman, Pauli, Heisenberg, and Landau are presented. Then the advent of the field of itinerant electron magnetism starting with the investigations of Bloch and Wigner, and more successful formulation by Slater and Stoner is presented.  The physical basis of the Slater-Stoner theory is discussed and its problems are summarized. Then, an overview of the debate between itinerant electron view of Stoner and localized electron view of Heisenberg is presented. Connected with this debate are the issues of exchange interactions.  The issues related to the origin of exchange interaction in Stoner model are discussed. We review the "middle-road" theory of van Vleck and Hurwitz--the very first theory which takes into account the electron correlation effects in the itinerant model. We close our presentation with the discussion of the very important issue of strong electron correlation in the itinerant picture.

\end{abstract}
}

\begin{document}

% Print the title
\maketitle

%----------------------------------------------------------------------------------------
%	ARTICLE CONTENTS
%----------------------------------------------------------------------------------------

\begin{figure}[!h]
\begin{center}
\includegraphics[height=5cm]{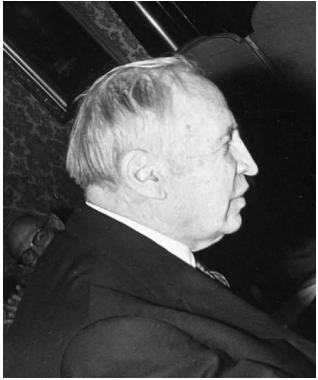}
\caption{Through this article we pay homage to John Hasbrouck Van Vleck (March 13, 1899 -- October 27, 1980) who set the foundation of theories of electron correlation with his "middle-road" theory.}
\end{center}
\end{figure}

This paper is divided into two parts: In the first part, an apparent paradox between the Langevin theory of paramagnetism and the Bohr - van Leeuwen theorem is presented and explained. Then, the problems in theoretical understanding of magnetism in the pre-quantum mechanical era (1900 - 1926) are presented. The  resolution of these problems started with the crucial contributions of van Vleck in the post quantum era (from 1926  to 1930s). Van Vleck's  key contributions are presented: (1) his detailed quantum statistical mechanical study of magnetism of real gases; (2) his pointing out the importance of the crystal fields or ligand fields in the magnetic behavior of iron group salts (the ligand field theory); and (3) his many contributions to the elucidation of exchange interactions in d electron metals. Next, the pioneering contributions (but lesser known) of Dorfman are discussed.  Then, in chronological order, the key contributions of Pauli, Heisenberg, and Landau are presented. Finally, the advent of the field of itinerant electron magnetism starting with the investigations of Bloch and Wigner, and more successful formulation by Slater and Stoner is presented.  The physical basis of the Slater-Stoner theory is discussed and its problems are summarized.

In the second part an overview of the debate between itinerant electron view (Stoner) and localized electron view (Heisenberg) is presented. Connected with this debate are the issues of exchange interactions.  These can be divided into two categories: (1) exchange interaction in itinerant models, and (2) exchange interaction in localized models. We start by discussing issues related to the origin of exchange interaction in Stoner model. Then we discuss the nature of exchange interaction in the Heisenberg model and an important working rule "the Slater curve" for the sign of this interaction. After highlighting its problems we introduce the contributions of Vonsovsky and Zener which introduce the idea of indirect s-d exchange interactions. Then Pauling's valence bond theory for the iron group metals is presented. Next comes the famous debate between the itinerant picture (Stoner model) and the localized picture (Heisenberg model).  Pros and cons of both approaches are discussed. The debate was settled in the favor of the itinerant model in the 1960s, when d-band Fermi surface was observed in iron group transition metals. However, the issue of correlation effects in the itinerant model remained open.  The debate still appears in its varied avatars in the current literature on unconventional superconducting strongly correlated materials.  Next, we briefly discuss the well settled issues of exchange interactions in insulator compounds (direct exchange; superexchange; and double exchange).  Then we review the "middle-road" theory of van Vleck and Hurwitz (the very first theory which takes into account the electron correlation effects in the itinerant model). We then introduce Friedel-Alexander-Anderson-Moriya theory of moment formation in pure iron group transition metals which is a kind of generalization of the famous Anderson impurity problem and further advances the "middle-road" ideas of Hurwitz and van Vleck. Finally the discussion of the very important issue of strong electron correlation in the itinerant picture is presented. 

\vspace{1cm}
{\huge{\centering{PART A}}}
\vspace{1cm}

\section{Failure of the classical picture: the Bohr-van Leeuwen theorem}
%—————————————————————————————————————————————————————————————————————————

The 19th century saw two major advancements in fundamental physics. One is the  "wedding" of electricity and magnetism through investigations of Oersted, Faraday, Maxwell and others.  The other major development occurred in the understanding of thermodynamical phenomena from molecular--kinetic point of view. Thermodynamical concepts like temperature, pressure, and thermodynamical laws were understood from the motion and interactions of atoms/molecule--the building blocks of matter. Maxwell for the first time used probabilistic or statistical arguments to derive the physical properties like pressure, viscosity etc of gases starting from the molecular--Kinetic point of view. This statistical method was greatly extended by Ludwig Boltzmann (and independently by Gibbs), and they transformed it into a well respected and highly successful branch of physics called statistical mechanics which bridged the gap between the microscopic dynamical laws that govern the motion of atoms and molecules and the macroscopic laws of thermodynamics.

One of the first successful application of statistical mechanics is the Langevin theory of paramagnetism (1905) [refer paper I\footnote{History of magnetism I: from Greeks to Paul Langevin and Pierre Weiss, Navinder Singh, hereafter referred as I.}]. However, there is one subtlety involved. In 1911, Niels Bohr in his PhD thesis applied the method of statistical mechanics to understand magnetism from atomic point of view. He concluded that within the setting of classical statistical mechanics it is not possible to explain any form of magnetism of matter! His method yielded zero magnetization. Thus there is an apparent contradiction between Bohr's approach and Langevin's approach, as both came in the pre-quantum era. 

The result of zero magnetism in classical statistical mechanics was re-discovered and elaborated independently in 1919 by Miss J. H. Van Leeuwen. The result is now famous as Bohr-van Leeuwen theorem. It can be explained in the following way\cite{1a}. Consider the case of a material in which all the degrees-of-freedom are in mutual thermodynamical equilibrium including electrons. In statistical mechanics thermodynamical quantities, including magnetization, are computed from free energy which can be expressed through partition function which is further expressed as a phase integral of the Boltzmann factor $(exp(-\frac{H}{k_B T}))$ involving the Hamiltonian ($H$). In an external magnetic field, the Hamiltonian ($\frac{p^2}{2m} + V(r)$) must be replaced by ($\frac{1}{2m} (p - \frac{e}{c}A)^2 +V(r)$) where $A$ is the vector potential and $p$ is the canonical momentum. It turns out that the phase integral (the partition function, $Z$) becomes independent of vector potential when the integration over momentum in the phase space integration is changed to $p'= p-\frac{e}{c}A$,  i.e., when momentum variable is changed. So the partition function becomes independent of vector potential, and resulting free energy  ($F = -k_B T ln Z$) also becomes independent of vector potential and magnetic field. It gives zero magnetization when differentiated ($M = -\frac{\pr F}{\pr H}$). In conclusion, this theorem raises an apparent paradox: how does magnetic effects arise in the Langevin theory which also uses classical statistical mechanics? Quantum mechanics was not known when Langevin advanced his theory (in 1905).

\section{Reconciling the Langevin theory with the Bohr-van Leeuwen theorem}
%————————————————————————————————————————————————————————————————————————

It turns out that the Langevin theory is {\it{not}} fully classical. It is actually semi-classical or semi-quantum in nature. Langevin did not consider all the degrees-of-freedom classically, as considered in the Bohr-van Leeuwen theorem. The internal motion of electrons within the atom which gives magnetism was not treated classically by Langevin.  He attributed a permanent magnetic moment to each atom without worrying about its origin. This state of affairs is best explained by J. H. Van Vleck\footnote{J. H. Van Vleck in  {\it Nobel lectures in physics 1971 - 1980}, Landquist (ed), World Scientific, 1992.}

"When Langevin assumed that the magnetic  moment of the atom or molecule had a fixed value $\mu$, he was quantizing the system without realizing it."

Assignment of a permanent magnetic moment to an atom is actually an introduction of  a quantum mechanical ingredient in to the problem which Langevin did not recognize explicitly. Also, Langevin did not take into account the space quantization (spin can only have discrete quantized values along the magnetization direction). In Langevin's theory magnetic moment can point in any direction and the phase integral was computed for all possible orientations.  Thus one can regard the Langevin theory as semi-classical, and the apparent paradox with {\it fully classical} Bohr-van Leeuwen theorem is immediately removed. As a side remark it is to be noted that when a fixed magnetic moment is assigned to an atom, one is departing from the principles of classical electrodynamics that an orbiting (i.e., accelerating) electron inside of an atom must radiate energy. Permanent magnetic moment implies permanent circling electrons inside the atom. The Langevin assumption of fixed magnetic moment directly leads to Bohr's principle of stationery states on which he built the quantum theory of the hydrogen atom.  However, Langevin did not explicitly state the stationarity of the circling electrons, and it was Bohr who fully recognized it, and stated it as an essential principle of the quantum theory\cite{2a}.

\section{Pre-quantum mechanical era and the problems of the old quantum theory}
%—————————————————————————————————————————————————————————————————————————

The success and failure of the old quantum theory of Bohr and others are well known\cite{2a}. And how the new quantum mechanics developed by Heisenberg, Born, Schroedinger, and Dirac replaced the patch-work of old quantum theory by a coherent picture of new quantum mechanics, in early 1920s, is also well known. In 1922, Stern-Gerlach experiment showed that magnetic moment of atoms can orient itself only in specific directions is space with respect to external magnetic field. This quantum mechanical phenomenon of spatial quantization was certainly missing in the Langevin treatment of paramagnetism. In the Langevin theory atomic moments can take any orientation in space. The required discretization of the spatial orientations was introduced, for the first time, by Pauli\footnote{Actually Pauli calculated electrical susceptibility. It turns out that same calculation goes through for magnetic susceptibility except one has to replace electric moment by magnetic moment\cite{3a}.} who found that susceptibility expression with respect to the temperature variation is the same as that of Langevin but with different numerical coefficient $C$ in $\chi = C\frac{N \mu^2}{k_B T}$. He found the value 1.54 instead of 1/3 of the Langevin theory.  Pauli used integer quantum numbers but analysis of the band spectrum showed the need for half-integer values. Linus Pauling revised Pauli's calculation by using half-integer instead of integer values, and it resulted in another value of the coefficient $C$\cite{3a}. The status of the field was far from satisfactory by 1925. There was another big problem. The calculations of susceptibility within the regime of old quantum theory appeared to violate the celebrated Bohr's correspondence principle, which states that in the asymptotic limit of high quantum numbers or high temperatures, the quantum expression should go over to the classical one ( as in black body radiation problem for $\frac{\hbar \om}{k_B T}<<1$). In the calculations of Pauli and Pauling there was no asymptotic connection with the Langevin theory. Then there was issues related to the weak and strong spatial quantization in the old quantum theory\cite{3a}. Also the origin of the Weiss molecular field remained a complete mystery. {\it In conclusion, the old quantum theory of magnetism was a dismal failure. }

\section{Quantum mechanical and post-quantum mechanical era, and the development of the quantum theory of magnetism}
%—————————————————————————————————————————————————————————————————————————————————————————————————

The modern quantum mechanics was in place by 1926. The equivalence of the matrix formulation of Heisenberg (1925) and wave-mechanical formulation of Schroedinger (1926) by shown by Schoredinger in 1926. In the same period van Vleck attacked the problem of magnetism with "new" quantum mechanics.

\section{Enter van Vleck}
%%%%%%%%%%%%%%%%%%%%%%%

One of the pioneer of the quantum theory of magnetism is van Vleck who  showed how new quantum mechanics could rectify the problems of the old quantum mechanics, and restored the factor of 1/3 of the Langevin's semi-classical theory. In doing so he took space quantization of magnetic moment into account (instead of the integral in the partition function, proper summation was performed). In one of the pioneer investigation, van Vleck undertook a detailed quantum mechanical study of the magnetic behavior of gas nitric oxide $(NO)$. He showed quantitative deviations from semi-classical Langevin theory in this case, and his results agreed very well with experiments\cite{4a}. The quantum mechanical method was applied to other gases, and he could quantitatively account for different susceptibility behavior of gases  like $O_2, NO_2$, and $NO$.\footnote{For a detailed account refer to his beautifully written book\cite{3a}.} The differences in magnetic behavior arise from the comparison of energy level spacings ($\hbar \om_{if}$) with the thermal energy $k_B T$. He showed that the quantum mechanical expression for susceptibility reduces to the semiclassical Langevin result when all energy level spacings are much less than the thermal energy ($|\hbar\om_{ij}|<<k_B T$).  In the opposite regime (when for all $|\hbar\om_{ij}|>>k_B T$ )  $\chi$ showed temperature independent behavior. In the intermediate regime ($|\hbar\om_{ij}|\sim k_B T$) susceptibility showed a complex behavior (the case of nitric oxide). Thus van Vleck re-derived the Langevin theory by properly taking into account the space quantization.

\begin{figure}[!h]
\begin{center}
\includegraphics[height=4cm]{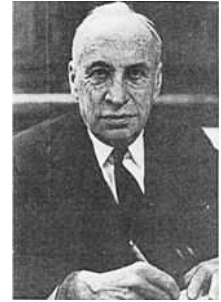}
\caption{Van Vleck (1899-1980). [Photo: Wikipedia Commons]}
\end{center}
\end{figure}

Another major contribution of van Vleck is related to magnetism in solid-state. When  a free atom (suppose a free iron atom) becomes a part in a large crystalline lattice (like iron oxide), its energy levels change. The change in the electronic structure of an atom is due to two factors (1) outer electrons participate in the chemical bond formation, thus their energy levels change, and (2) in a crystalline lattice, the remaining unpaired electrons in the outer shells of an atom are not in a free environment, rather they are acted upon by an electrostatic field due to electrons on neighboring atoms. This field is called the crystalline field.     

Van Vleck and his collaborators introduced crystalline field theory (also known as the ligand field theory in chemical physics departments) to understand magnetic behavior in solid-state. With crystalline field ideas they could understand different magnetic behaviors of rare earth salts and iron group salts. It turns out that in rare earth salts $4f$ electrons are sequestered in the interior of the atom, and do not experience the crystalline field very strongly (figure 2). The energy level splitting due to crystalline electric field is small as compared to thermal energy ($k_B T$), and it remains small even at room temperatures. Due to this the magnetic moment of the atoms behave as if the atom is free and shows the Langevin-Curie behavior $\chi\sim\frac{1}{T}$\cite{4a,5a,6a}.

\begin{figure}[!h]
\begin{center}
\includegraphics[height=3cm]{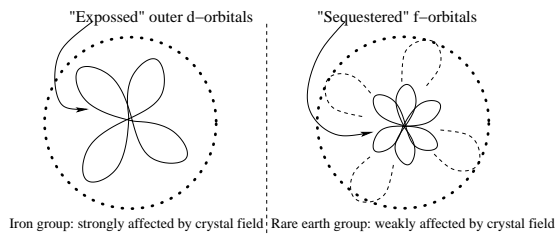}
\caption{A cartoon showing why crystal field effects differently an iron group ion and a rare earth ion.}
\end{center}
\end{figure}

In contrast to this case, in the iron group salts crystalline field is so strong that it quenches a large part of the orbital magnetic moment, even at room temperatures, leaving mainly the spin part to contribute to magnetism of salts of iron (refer to figure 2). 

Magnetism of iron group {\it metals} is a different story (as compared to salts). In this case it turns out that charge carriers are also responsible for magnetism. The magnetism due to {\it itinerant electrons} was developed by Bloch, Slater, and Stoner (refer to part B). The other extreme of localized electrons was investigated by Heisenberg. Van Vleck advanced ideas that can be dubbed as "middle of the way" approach (refer to part B). For his pioneering contributions van Vleck was awarded with the Nobel  prize in physics in 1977 along with Phil Anderson and Nevill Mott. His articles are beautifully written and extremely readable and should form an essential element in a course (graduate or undergraduate) on magnetism. One can say that van Vleck is the father of the modern theory of magnetism, and his name will be forever remembered.

\section{Enter Dorfman}

When quantum mechanical study of magnetism of real gases was started by van Vleck in mid 1920s, the quantum mechanical study of magnetism in metals also started in the other continent transatlantic. 

The discovery of the paramagnetic properties of conduction electrons in metals is generally attached to Wolfgang Pauli. Pauli's paper came in 1927. Even before that, in 1923, Russian physicist Yakov Grigor'evich Dorfman (figure 3) put forward the idea that conduction electrons in metals posses paramagnetic properties\cite{7a}. His proposal was based on a subtle observation: when one compares susceptibility of a {\it diamagnetic} metal with its ion, the susceptibility of the ion is always greater than its corresponding metal. It implies that there is some positive susceptibility in the case of the diamagnetic metal that partly cancels out the larger negative diamagnetic susceptibility. And this cancellation is prohibited in the case of metal's ion (due to ionic bonding). It was Dorfman's intuition that some positive susceptibility is to be attributed to conduction electrons in the metal i.e., some paramagnetic susceptibility has to be there. \footnote{It is important to note that the notion of the electron spin came in 1925 with a proposal by Uhlenbeck and Goudsmit and paramagnetism due to electron spin was discovered in 1927 by Pauli as mentioned before. But Dorfman's proposal came in 1923!} Dorfman's conclusion is based on his careful examination of the experimental data. After the discovery of the electron spin, Pauli gave the theory of paramagnetism in metals due to free electron spin. However, Dorfman was the first to point out paramagnetism in metals\cite{7a}.

\begin{figure}[!h]
\begin{center}
\includegraphics[height=4cm]{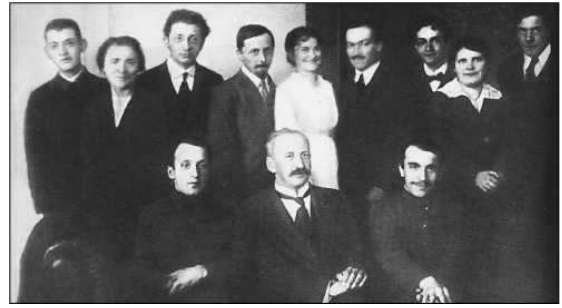}
\caption{Yakov Grigor'evich Dorfman (1898-1974) standing on extreme left. The person sitting in the center is  A .F. Ioffe.  [Photo: Wikipedia Commons]}
\end{center}
\end{figure}

One of the other important contributions of Dorfman is his experimental determination of the nature of Weiss molecular field responsible for ferromagnetism in the Weiss theory. It was believed that the Weiss field is of magnetic origin due to which spins align to give a net spontaneous magnetization. To determine whether the Weiss field is of magnetic origin or of non-magnetic origin, Dorfman passed beta-rays (a free electron beam) in two samples of nickel foils, one magnetized and the other unmagnetized. From deflection measurements he determined that Weiss field is of non-magnetic origin\cite{8a}.

In conclusion, Dorfman was an early contributor to the quantum theory of magnetism. But he is not  as well known as he should have been. 

\section{Enter Pauli}

\begin{figure}[!h]
\begin{center}
\includegraphics[height=4cm]{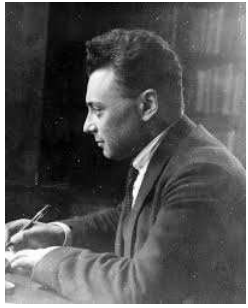}
\caption{Wolfgang Pauli (1900 - 1958). [Photo: Wikipedia Commons]}
\end{center}
\end{figure}

Pauli's contribution to magnetism is well known. He formulated paramagnetic behavior of conduction electrons in metals in 1927 and showed that paramagnetic susceptibility is temperature independent (in the leading order). The derivation is discussed in almost all books devoted to magnetism and solid state physics\cite{9a}. Pauli's derivation of the paramagnetic susceptibility can be described as one of the early application of Fermi-Dirac statistics of electrons in metals. In the standard derivation\cite{9a} one calculates the thermodynamical potential $\Omega (H)$ of free electron gas in a magnetic field $H$.  Magnetization is obtained by the standard algorithm of statistical mechanics: $M = -\frac{\pr \Omega}{\pr H}$, and susceptibility $\chi = \frac{\pr M}{\pr H}$. For illustration purpose there is a simpler argument\cite{10a} which goes like this. For metals at ordinary temperatures one has $k_B T<<E_F$ where $T$ is the temperature and $E_F$ is the Fermi energy. Thus electrons only in a tiny {\it diffusion zone} around the Fermi surface participate in thermodynamical, electrical, and magnetic properties (other electrons are paired thus dead). If $N$ is the total number of electrons, then fraction of electrons in the diffusion zone is $N\frac{T}{T_F}$ where $T_F$ is the Fermi temperature ($k_B T_F = E_F$).  Each electron in the diffusion zone has magnetic susceptibility roughly given by $\chi \sim \frac{\mu^2}{k_B T}$ where $\mu$ is its magnetic moment. Thus total magnetic susceptibility of metal is given by:  $N\frac{T}{T_F} \times \frac{\mu^2}{k_B T} = N\frac{\mu^2}{k_B T_F}$ which is independent of temperature as the more accurate calculation shows.

\section{Enter Heisenberg}

\begin{figure}[!h]
\begin{center}
\includegraphics[height=4cm]{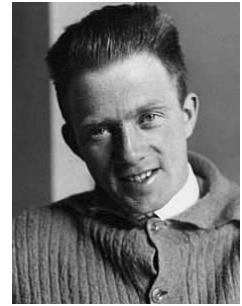}
\caption{Werner Heisenberg (1901 - 1976). [Photo: Wikipedia Commons]}
\end{center}
\end{figure}

As mentioned before Dorfman in 1927 pointed out that the Weiss molecular field required in the theory of ferromagnetism is of non-magnetic origin. The puzzle of the Weiss molecular field was resolved by Heisenberg in 1928. The central idea is that it is the {\it quantum mechanical exchange interaction} which is responsible for the ferromagnetic alignment of spins. Quantum mechanical exchange interaction has no classical analogue, and it results due to the overlapping of orbital wave functions of two nearby atoms. Symmetry of the hybrid orbital is dictated by the nature of the spin alignment which obeys the Pauli exclusion principle. Thus there is an apparent spin-spin coupling due to orbital symmetry and under specific circumstances the ferromagnetic spin alignment significantly lowers the bonding energy thereby leading to a stable configuration.\footnote{It is very important to note that energy associated with spin-spin coupling of two electrons via exchange is very  large as compared to the magnetic dipole-dipole interaction energy which is given by \[V_{ij} = \frac{u_i.u_j}{r_{ij}^3} - 3 \frac{(u_i.r_{ij})(u_j.r_{ij})}{r_{ij}^5}.\] This very small magnetic energy cannot lead to ferromagnetic alignment. In other systems, like ferro-electrics it is an important energy.} 

The Heisenberg model based on exchange interactions is related to the resonance-energy-lowering model for chemical bonding by Heitler and London\cite{11a}.  In the Heitler-London theory of the chemical bond in hydrogen molecule, it is the exchange of electrons on two hydrogen atoms that leads to the resonant lowering of the energy of the molecule. Electrons stay in an antiparallel spin configuration  thereby enhancing the overlap of orbital wave functions in the intermediate region of two hydrogen atoms. This leads to bond formation.  This idea of resonant lowering of energy via exchange of electrons is greatly used by Linus Pauling in his general theory of the chemical bond\cite{11a}. The Heisenberg model is built on similar ideas and goes like this\cite{5a,12a}. Let $S_i$ be the total spin at an atomic site $i$. If exchange interaction between nearest neighbors is the only one important, then the interaction energy (under certain approximations\footnote{Here $S_i$ is the total spin at an atomic site $"i"$, i.e., it includes a vector sum over all the spins of unpaired electrons. In our notation $i$ and $j$ label two nearest sites.   Let $m$ and $n$ denote orbital numbers on a given site $i$ or $j$ (in cases where there are many unpaired spins in different orbitals).  Exchange interaction energy between an electron in $m$th orbital at site $i$ and an electron in $n$th orbital at site $j$ is given by \[V_{i,m;j,n} = -2 J_{i,m;j,n} S_{i,m} . S_{j,n}.\]  Total interaction is obtained by summing over all $m$ and $n$ \[V_{i,j} = -2 \sum_{m,n} J_{i,m;j,n} S_{i,m} . S_{j,n}.\] The main assumption is that the exchange integral between  $m$th orbital at site $i$ and $n$th orbital at site $j$ is {\it assumed to be independent of $m$ and $n$. It is like assuming the same exchange integral between two $s$-orbitals or two $d$-orbitals or between  $s$ and $d$ orbitals on two different sites $i$ and $j$. That is}\[J_{i,m;j,n}\simeq J_{i,j} \simeq J.\]  Validity of this assumption depends crucially on the nature of the system under consideration.  Of course, overlap of two S-orbitals is different  from that of two d-orbitals. But let us accept this assumption. Under this assumption $V_{i,j} = -2 J_{ij}S_i.S_j$ where $S_i = \sum_n S_{i,n}$ etc. Hence one obtains the Heisenberg model as given in the main text.}) is given by

\[V_{ij} = -2 J_{ij} S_i.S_j.\]

$J_{ij}$ is called the exchange integral\footnote{\[J_{ij} = \int d\tau_1\int d\tau_2 \phi_i(1)\phi_j(2) H_c \phi_j(1)\phi_i(2).\]}.  For ferromagnetism the sign of $J_{ij}$ has to be positive, and for anti-ferromagnetism it has to be negative. The question on what parameters the sign of $J$ depends is complicated and vexed one (we will discuss these issues in part B).

The above exchange interaction is now known as the Heisenberg exchange interaction or the direct exchange interaction. There is a variety of exchange interactions (both in metals and insulators) that will be discussed in part B.   

To compare predictions of the model with experiment, one needs its solution. The very first solution provided by Heisenberg himself is based on some very restrictive assumptions.  So tight agreement with experiments may not be expected, and it leads to some qualitative results. Heisenberg used complicated group theoretical methods and a Gaussian approximation of the distribution of energy levels to find an approximate solution.\footnote{An alternative and comparatively simpler method was provided by Dirac using the vector model with similar conclusions\cite{1a,2a}.} From his solution Heisenberg observed that ferromagnetism is possible only if the number of nearest neighbors are greater than or equal to eight ($z=8$). This conclusion is certainly violated as many alloys show ferromagnetism with $z=6$. The second result which is much more important is that of magnitude of $\lambda$ it turns out that $\lambda$ of the Weiss molecular field takes the form

\[\lambda = z\frac{J}{2 N \mu_B^2}.\]

{\it The large value of $\lambda$ required for ferromagnetism is not a problem anymore, as the exchange integral $J$ can be large, thus resolving the problem of Weiss theory. This is the biggest success of the Heisenberg model. }

In conclusion, Heisenberg's model resolved the puzzle of the Weiss molecular field using the concept of exchange interaction. This concept turns out to be the key to the modern understanding of magnetism in more complex systems. Heisenberg's solution was based on many drastic assumptions which were later improved upon. Literature on the Heisenberg model and its various approximate solutions is very vast. Some references are collected here\cite{5a,6a,12a,13a}.

%\section{Attempts to solve the Heisenberg hamiltonian}

\section{Enter Landau}

Metals which are not ferromagnetic show two weak forms of magnetism, namely, paramagnetism and diamagnetism. Paramagnetism we have discussed, diamagnetism due to free conduction electrons is a subtle phenomenon and was a surprise to the scientific community\cite{1a} when Lev landau discovered it in 1930.  To appreciate it consider the following example. Consider the classical model of an atom in which a negatively charged electron circulates around a positive nucleus. A magnetic moment  will be associated with the circulating electron (current multiplied by area).  Let  a uniform magnetic field be applied perpendicular to the electrons orbit. Let the magnitude  of the magnetic field be increased from zero to some finite value. Then, it is an easy exercise  in electrodynamics to show that an electromotive force will act on the electron in such a manner that will try to oppose the increase in the external magnetic field (i.e., Lenz's law). The induced opposing current leads to an induced magnetic moment in the opposite direction to that of the external magnetic field, and the system shows a diamagnetic behavior (induced magnetization in the opposite direction to the applied magnetic field). 

However, when a collection of  such classical model-atoms is considered the diamagnetic effect vanishes. The net peripheral current from internal current loops just cancels with the opposite current from the skipping orbits (refer, for example, to \cite{1a}). This observation also agrees with the Bohr-van Leeuwen theorem of no magnetism in a classical setting. Thus in a classical setting it is not possible to explain the diamagnetic effect.

%\begin{figure}[!h]
%\begin{center}
%\includegraphics[height=4cm]{Landau.eps}
%\caption{Lev Landau (1908 - 1968). [Photo: Wikipedia Commons]}
%\end{center}
%\end{figure}

However, in 1930, Landau surprised the scientific community by showing that free electrons show diamagnetism which arises from a quantum mechanical energy spectrum of electrons in a magnetic field. As described in many text books\cite{9a} the solution of the Schroedinger equation for a free electron in a magnetic field is similar to that of the solution of the harmonic oscillator problem. There exits  equally  spaced energy levels - known as Landau levels. Each Landau level has macroscopic degeneracy. Statistical mechanical calculation using these Landau levels shows that there is non-zero diamagnetic susceptibility associated with free electrons which is also temperature independent as Pauli paramagnetism is.  And as is well known Landau level physics plays a crucial role in de Haas - van Alphen effect and related oscillatory phenomena, and in quantum Hall effects.

\section{Enter Bloch and Wigner}
%\begin{figure}[!h]
%\begin{center}
%\includegraphics[height=4cm]{bloch.eps}
%\caption{Felix Bloch (1905 - 1983). [Photo: Wikipedia Commons]}
%\end{center}
%\end{figure}
Bloch in 1929\cite{14a} advanced the idea that magnetism in iron group metals might be originating from itinerant electrons (in contrast to Heisenberg's localized electron model)\footnote{The idea that itinerant electrons might be responsible for ferromagnetism was already there. Frenkel in 1928 discussed the possibility of ferromagnetism due to itinerant electrons via Hund's coupling. Experimental investigations were made by Dorfman, Kikoin, and their colleagues\cite{7a}.}. The basic principle behind Bloch's theory is as follows. As is well known conduction electrons form a sphere in momentum space known as the Fermi sphere. Each momentum state is doubly occupied with one electron of up spin and the other with down spin. This configuration minimizes  the total kinetic energy (K.E.) of the system. Now, if there is an exchange interaction between the conduction electrons then they tend to align their spins. Pauli's exclusion principle then prohibits them to be in the same momentum state, and electrons must migrate to higher momentum states. This migration of electrons to higher momentum states leads to increased K.E. of the system. Thus there is a competition between exchange interaction energy which tends to lower the energy of the system by aligning spins of electrons and K.E. which tend to pair them up with two electrons in each momentum state. Under "suitable conditions" exchange interaction is the winner and system becomes unstable to ferromagnetism.  The "suitable conditions" according to Bloch are sufficiently low electron density or sufficiently large  electron mass (this will be made more precise in the following paragraph). But Bloch's argument has problems as was first pointed out by Wigner. Before we present Wigner's argument, let us discuss  exchange in Bloch's picture in little more detail. 

Exchange energy  basically originates from electrostatic Coulomb repulsion between two electrons. By having parallel spins two electron are spatially pulled apart due to Pauli's exclusion principle, and this lowers the electrostatic energy of the system.  Bloch showed that when

\[4.5 \frac{e^2 m^*}{h^2}> n^{1/3}\]

the system exhibits ferromagnetism. That is when, either, electron density  ($n$) is very low, or, when effective mass of electron is very large.

\begin{figure}[!h]
\begin{center}
\includegraphics[height=4cm]{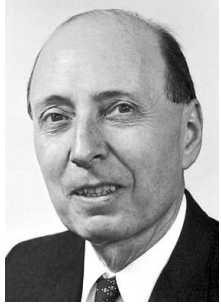}
\caption{Eugene Wigner (1902 - 1995). [Photo: Wikipedia Commons]}
\end{center}
\end{figure}

The above condition is too restrictive. Wigner\cite{15a}  in 1938 showed that Bloch's argument is not realistic one in that it neglects Coulomb electrostatic interaction between electrons with anti-parallel spins. This "correlation energy" is not taken into account in Bloch's calculation. Wigner estimated these correlation effects and showed that the possibility of ferromagnetism in Bloch's picture is nil.

\section{ Enter Slater and Stoner and the arrival of the itinerant electron magnetism}
%
%\begin{figure}[!h]
%\begin{center}
%\includegraphics[height=4cm]{slater}
%\caption{John C. Slater (1900 - 1976). [Photo: Wikipedia Commons]}
%\end{center}
%\end{figure} 

Slater in 1936\cite{16a} discussed the possibility of ferromagnetism due to itinerant electrons. He argued that the exchange interaction responsible for the spin alignment of itinerant electrons is not the itinerant exchange, as argued by Bloch, but it is of the {\it intra-atomic origin}\cite{17a}. It is an extension of the Hund rule of maximum total spin for a less than half filled shell\footnote{Similar ideas were advanced by Frenkel as mentioned before in a previous footnote.}.  Consider the case of itinerant electrons of narrow d-band in iron group metals. An itinerant electron flits from one atom to another thereby changing atom's polarity (that is atomic ionization states change and it is also known as the polar model in contrast to Heisenberg's non-polar model where the polarity of an atom remains the same due to localized electrons).  The minimum energy configuration is that when this itinerant electron has the same spin polarization as that of electrons already there in the corresponding shell of the atom with degenerate orbitals.  When this electron flits from that atom to a nearby one it takes with it its prejudice of being in that spin configuration. For example, if electrons in less than half full degenerate d-orbitals of an atom have spin polarization along the positive z-direction (say) then this flitting electron will have its spin polarized along the same direction.  In the nearby atom the very same mechanism works and it leads to spin alignment. In total, this intra-atomic exchange leads to ferromagnetic state.\footnote{Van Vleck also developed a model on similar lines called the "minimum polarity model" which is discussed in Part B and references to the related literature are given there.} In the language used here we are using both "band" concept and "orbital" concept at the same time. It appears incoherent, but it turns out that valence electrons of the iron group elements retain their atomic character to some extent\cite{6a}. 

With the Slater model one can appreciate the fact that alkali and alkaline earth metals are not ferromagnetic as in these metals intra-atomic exchange is not possible in the conduction s-bands as these are non-degenerate. On the other hand d-band metals can be ferromagnetic as intra-atomic exchange can provide the required mechanism for spin alignment due to d-band degeneracy. 

But this criterion based on degenerate versus non-degenerate bands leaves open the question of no ferromagnetism in p-band metals which are also degenerate.\footnote{The complications due to various exchange interactions are discussed in detail in part B.}

In contrast to all these complications Stoner in 1936 adopted a completely phenomenological approach\cite{18a}.  He basically superposed Weiss molecular field (i.e., the exchange field) on itinerant electrons without worrying much about the origin of the exchange interaction in metals. The stoner theory is computationally successful and its results can be compared with experiments. The basic mechanism of ferromagnetism in the Stoner theory is the same as that of Bloch's----the competition between exchange energy and the K.E. At zero  temperature this  leads to the following condition for ferromagnetism:

\[ I \rho(E_F)>1.\] 

Here $I$ is the average exchange interaction energy and $\rho(E_F)$ is the electronic density of states (EDOS) at the Fermi level. So, according to the Stoner condition, metals having large value of EDOS at the Fermi level or having large value of exchange interaction are tend to be ferromagnetic. For example d-band metals have a chance of being ferromagnetic as EDOS for d-band is large, whereas EDOS for s-band and p-band metals is smaller and they are not ferromagnetic. Not all d-bands metals are ferromagnetic, so Stoner model is definitely not the complete answer. But it captures the phenomenon in a qualitative way.

\begin{figure}[!h]
\begin{center}
\includegraphics[height=4cm]{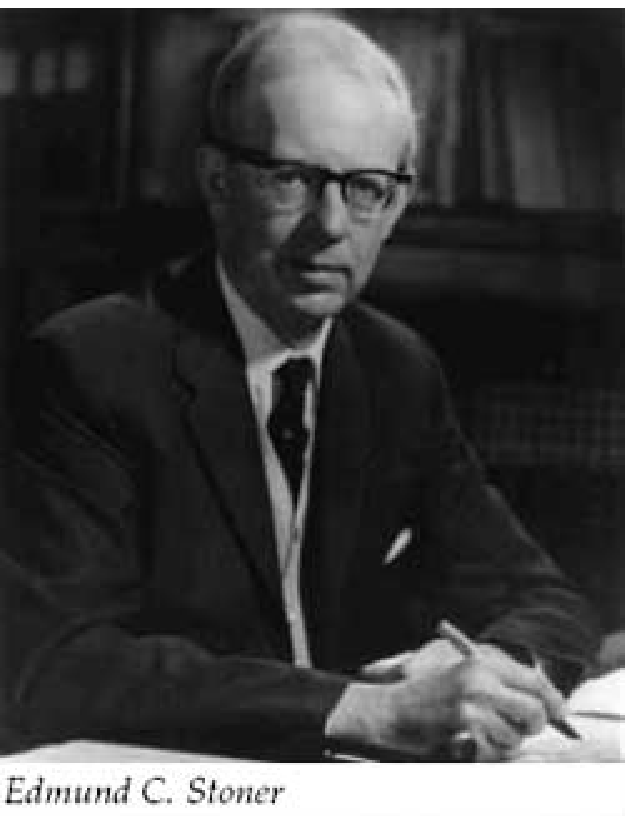}
\caption{Edmund C. Stoner (1900 - 1976). [Photo: Wikipedia Commons]}
\end{center}
\end{figure} 

Stoner also performed extensive calculations of temperature dependence of ferromagnetism. The finite temperature model can be easily described in the following way. Let $\Delta$ be the energy due to internal exchange field which is given by $I M$ where $M$ is the uniform magnetization (in the literature $\Delta$ is also called the band splitting). Let $N$ be the total number of electrons given by

\[N = \int_{-\Delta}^\infty d\ep \rho(\ep + \Delta) f(\ep) +\int_{\Delta}^\infty d\ep \rho(\ep - \Delta) f(\ep),\] 

where $f(\ep) = \frac{1}{e^{\beta(\ep-\mu)} + 1}$ and $\Delta = I M$. And the total magnetization is given by

\[M = \frac{1}{2}\int_{-\Delta}^\infty d\ep \rho(\ep + \Delta) f(\ep)  - \frac{1}{2}\int_{\Delta}^\infty d\ep \rho(\ep - \Delta) f(\ep).\] 

These two equations can be easily solved numerically in a self-consistent manner if the EDOS is given as a function of energy. When magnetization is plotted as a function of temperature in the case of parabolic band one obtains $M-T$ graph which roughly agrees with that obtained with the Weiss theory. Actually, in the limit $\frac{I}{E_F}>>1$ Stoner's $M-T$ curve exactly matches with that of Weiss localized model with $S=1/2$ on each site. This agreement is expected. When exchange energy is much greater than Fermi energy then width of the band is negligible as compared to exchange energy scale, and then the results of a localized model are expected to appear.

\begin{figure}[!h]
\begin{center}
\includegraphics[height=4cm]{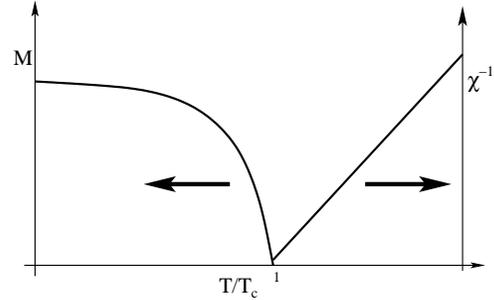}
\caption{Magnetization and inverse susceptibility as a function of scaled temperature in Stoner theory.}
\end{center}
\end{figure} 

However, the plot of $\frac{1}{\chi}$ i.e., inverse susceptibility versus temperature shows significantly more curvature as compared to the well obeyed Curie-Weiss law (which is a straight line for inverse susceptibility versus temperature). This is a drawback of the Stoner model. The other major drawback of Stoner theory is that when $T_c$ is calculated using magnitude of the saturation magnetization it results in a very high values of $T_c$. Sometimes even an order of magnitude larger\cite{19a}. Thus, one can say that Stoner model is not quantitatively successful, but qualitatively it captures the phenomenon of itinerant electron magnetization. The story how Random Phase Approximation (RPA) and the Moriya-Kawabata theory improves upon it is presented in  next article of the current series.

\section{Physical basis of the Slater-Stoner theory}

As discussed in the previous section the Slater model can provide a basis to understand magnetism of iron group metals using the idea of d-band degeneracy. The intra-atomic exchange can provide the required spin alignment via an extension of Hund's mechanism between a flitting electron of d-band and its localized companion in other d-orbital. On the other hand missing orbital degeneracy in the valence orbitals in alkali and alkaline earth metals blocks this mechanism of Slater, and hence these turn out to be non-ferromagnetic.  But this mechanism of Slater leaves open the question of why no ferromagnetism in p-band metals which are also degenerate.

The issues are partly resolved by using Stoner theory.  Stoner condition requires that for having favorable circumstances for ferromagnetism in a metal, there should be large exchange interaction energy or large value of EDOS. EDOS for s-band and p-band metals is smaller as compared to that in d-band metals. But as mentioned before not all d-bands metals are ferromagnetic.  Thus both approaches have their own problems. Although Stoner theory provides a clue and quantitative results but a complete answer cannot be given within Slater-Stoner ideas. Many other exchange interaction ideas were advanced which are discussed in PART B. Ferromagnetism in d-band metals is a complicated issue, and still we do not have full understanding!

%%%%%%%%%%%%%%%%%%%%
\vspace{1cm}
{\huge{\centering{PART B}}}
\vspace{1cm}

\section{Heisenberg versus Stoner}

In Part A, the Heitler-London approach motivated Heisenberg model and the Bloch-Wigner-Slater approach motivated Stoner model are discussed. In the Heisenberg model, electrons which are responsible for ferromagnetism are localized on atomic sites. The localized electron picture is true for magnetic insulator compounds, but it is not true for metals in which charge carriers are also responsible for magnetic effects (as in the case of 3d transition iron group metals). There are experimental proofs of it, for example, ferromagnetic transition metals show large electronic specific heat, and d-electron Fermi surfaces\cite{6a}.

In the opposite picture of itinerant electrons, contributions of Bloch, Wigner, Slater, and Stoner are discussed in part A. In a nutshell, Stoner superposed Weiss molecular field (or exchange field) on the Sommerfeld free electron model of metals. Free electrons undergo spin polarization under the action of Weiss molecular field or exchange field which leads to ferromagnetism. The finite temperature behavior of the Stoner model can be studied using standard method of statistical mechanics i.e., by calculating the free energy etc. However, the Stoner model alone is not sufficient to understand magnetic properties of iron group metals and Heisenberg model alone is not sufficient to understand magnetic properties of insulating systems, as will be discussed in subsequent sections.

The above picture was not available before early 1950s. So it was not clear whether Stoner model is more appropriate or the Heisenberg model to discuss ferromagnetism of iron group metals. Central to this dichotomy was the question whether $d$ electrons in iron group metals are localized or itinerant (de Haas - van Alphen Fermi surface studies of transition metals came in the late 1950s). Thus at that time it was a real confusion whether Heisenberg model is more appropriate for ferromagnetic transition metals as it reproduced the experimental Curie-Weiss law very well, or, the itinerant Stoner model as it reproduced fractional Bohr magneton numbers of saturation magnetization.  Heisenberg model failed to reproduce fractional magneton numbers while itinerant  Stoner model failed to reproduce the Curie-Weiss law\footnote{Stoner model leads to much more curvature in the graph of inverse susceptibility versus temperature whereas it should be linear according to the Curie-Weiss law.}.  Before we enter into this very interesting debate and list pros and cons of both models, we would like to delve into much more important and deeper questions related to the origin of Weiss molecular field in the itinerant model and the origin and sign of exchange interaction $J$ in the Heisenberg model. Answers to these questions help to understand and resolve the debate.

\section{The issue of the origin of Weiss field in the Stoner model}

At a more fundamental level, the origin of the Weiss molecular field in the itinerant model was attributed to intra-atomic exchange (within an atom) by Slater. In other words, Slater's intra-atomic exchange mechanism provides a microscopic basis to the Hund rules which states that if d shell in a transition metal ion is less than half full then spins tend to align parallel to each other to give maximum total S. In this way phenomenologically introduced Weiss molecular field in the Stoner model receives its microscopic justification. But the intra-atomic exchange mechanism has its own problems as was discussed in part A. The other possible explanation was given by Bloch. The exchange interaction between free electrons is "inherently" ferromagnetic. But, it to be effective, electronic density has to be low to make exchange energy dominant over kinetic energy (refer to part A). This condition is not valid for transition metals. On the top of it, Wigner pointed out that electronic correlation effects completely destroy the effect of exchange interactions. Thus, Slater's intra-atomic exchange is the most likely candidate.

\section{The issue of the sign of J in the Heisenberg model}

For occurrence of ferromagnetism in the Heisenberg model

\[H = -J \sum_{<ij>} S_i.S_j\]

the sign of $J$ must be positive. The Heisenberg model is motivated by the homopolar bond formation theory of Heitler and London. In the bond formation, say in hydrogen molecule, it is the exchange of electrons that leads to resonant lowering of energy.  Here the exchange interaction turns out to be negative, and electrons pair up in the hybrid molecular orbital with anti-parallel spins whereas in the Heisenberg model for ferromagnetism exchange interaction has to be positive and electrons must have parallel spins to show ferromagnetism. Thus it seems difficult to reconcile two opposite pictures: one requiring positive $J$ for ferromagnetism (Heisenberg), and the other requiring negative $J$ for chemical bond formation (Heitler-London) whereas both originate from exchange mechanism. 

In his original contribution (refer to \cite{2} page 192) Heisenberg argued that the sign of $J$ is positive in ferromagnetic metals because large principle quantum numbers are involved in this case whereas in chemical bond problems principle quantum numbers involved are smaller.  With this idea one can reconcile the opposite pictures. However, Heisenberg's guess is wrong as according to his argument metals in 2d and 3d transition periods of the periodic table have still larger principle quantum numbers, but they are not ferromagnetic.   

This situation was made partly clear by a much more relevant argument due to Slater\cite{12a}. He argued that positive sign of $J$  in ferromagnetic metals should be attributed to larger interatomic distances as compared to atomic radii involved. And the sign of $J$ changes from negative to positive when interatomic distance is varied from smaller to larger value with respect to the atomic radius.

\begin{figure}[!h]
\begin{center}
\includegraphics[height=3.5cm]{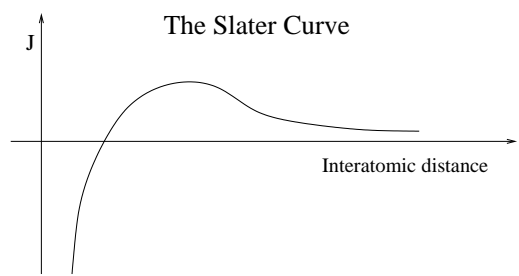}
\caption{The Slater curve.}
\end{center}
\end{figure}

So called "the Slater curve" (depicted in the above figure) not only explains why ferromagnetism does not occur in the second and third row transition elements (as inter-atomic distance is too small) but also why only last elements of the first row show ferromagnetism (as interatomic distance is just appropriate). Thus the Slater curve provided a "rule of thumb" when to expect $J$ to be positive. However, Slater's idea is also not free from criticism.  There is no single example where it is theoretically proved that $J$ in a given ferromagnetic material is positive.  Realistic theoretical calculations to compute J are extremely complicated as wave functions deform from free atomic state to something complicated when it is present in a matrix.\footnote{A computation of \[J= \int d\tau_1\int d\tau_2 \phi_i(1)\phi_j(2) V \phi_j(1)\phi_i(2)\] requires a thorough knowledge of wavefunctions which are not exactly known in an environment where atom is present in condensed state.} Thus Slater's guess remained unproved (i.e., without theoretical justification, although empirically it seemed possible) and further investigations were needed. 

Keeping roughly the chronological order, we next discuss the  Zener-Vonsovsky model which leads to positive $J$ through a different mechanism.

\section{Enter Vonsovsky and Zener}

We state at the outset that Vonsovsky-Zener model as originally invented for ferromagnetic d-electron metals is not the correct mechanism responsible for ferromagnetism in transition metals. Vonsovsky and zener maintained, when they put forward the idea\footnote{The basic idea is due to S. V. Vonsovskii 1946\cite{4}, and later on developed by Zener in 1951\cite{5}.},  that d-electrons form an isolated systems with localized electrons while s-electrons form running waves i.e., bands. This was clearly in contradiction to later experimental investigations using de Haas van Alphen effect  which showed d-electron Fermi surfaces in the early 1960s.  So d-electrons are itinerant rather than localized. Therefore  Vonsovsky-Zener (VZ) model were to be discarded.  However, these ideas form seeds of very important progress  in understanding magnetism in f-electron systems and dilute magnetic alloys in which f-electrons can be treated localized and which further lead to the development of the Kondo effect and RKKY (Ruderman-Kittel-Kasuya-Yosida)  interaction. We will not delve into these very interesting topics. There is a vast literature on these, and  interested reader can consult\cite{}. Here we present the debate regarding the sign of $J$ in the Heisenberg model, and how Vonsovsky-Zener (VZ) model leads to a positive $J$ through an entirely different mechanism. 

The VZ mechanism states that there is a Hund's type coupling of highest multiplicity between conduction s-electrons and localized d-electrons in partly filled d-shells\cite{6a}. To imagine this mechanism one can consider this picture. Consider that a flitting s-electron enters into a partly filled d shell. It stays there for a tiny time interval of the order of $\frac{\hbar}{E_{ion}}$ where $E_{ion}$ is the energy required to remove that electron from partly filled d shell, i.e., ionization energy. During this tiny time interval (which is of the order of femtoseconds) Hund's mechanism works and it tends to align its spin in the same direction as that of already present d electrons. Thus there is an effective ferromagnetic coupling between s electron spin and d- electron spin, and VZ model postulate that it can be written as $-\beta S_d S_s$ where $\beta$ is a positive parameter of the model and $S_d$ ($S_s$) is the total spin of d (s) electrons. The internal interactions between two $s$ electrons, and between two $d$ electrons is assumed to be anti-ferromagnetic\footnote{The internal exchange interaction between two conduction s electrons is ferromagnetic in nature as first pointed out by Bloch. However, this to be effective required very low electron density (part A). This is not the case with ferromagnetic metals and kinetic energy wins over exchange energy and electrons pair up with anti-parallel spins in a given momentum state. Also Wigner's correlation effects completely destroy parallel alignment in Bloch's model. The exchange interaction between localized d electrons is assumed to be anti-ferromagnetic.}.  Thus the total interaction energy can be written as

\[E = \frac{1}{2} J S_d^2 + \frac{1}{2}\gamma S_s^2 - \beta S_d S_s.\]

For a given value of $S_d$, energy is minimized when $S_s = \frac{\beta}{\gamma} S_d$ and one can write that $E = -\frac{1}{2} J_{eff} S_d^2$ where

\[J_{eff} =  \frac{\beta^2}{\gamma} - J.\]

Thus there is an effective interaction between $d$ electrons with exchange coupling $J_{eff}$. This can be clearly positive if $\beta$ is sufficiently large ( that is coupling between s electrons and d electrons is sufficiently large). Zener did quantitative calculations to prove his point and got partial success\cite{1,4}. Later on refined calculations showed that induced interaction between d-electrons via conduction $s$ electrons is not exactly ferromagnetic but it has a complex oscillatory character as a function of distance between $d$ electrons (i.e., RKKY interaction). Thus VZ model was an oversimplified model and had to be abandoned.

It turns out that Heisenberg model alone in its original formulation cannot address the issue of ferromagnetism in iron group metals. It is applicable to a few systems like $CrO_2$ and $CrBr_3$\cite{6}. And its advanced versions can be applied to a variety of magnetic insulator compounds. Below we present another "failed theory" of ferromagnetism in iron group metals. This has historical value only, however, it leads to other important concepts and developments.

\section{Enter Pauling with his valence bond theory}

Linus Pauling advanced a theory of ferromagnetism in iron group metals in 1953\cite{1,7}. His theory is an application of his resonating valence bond ideas which are successful in the chemical bond theory. But these ideas were not very successful when applied to metals. Central to his theory is the concept of hybridization. He explained the ferromagnetism of iron group metals in the following way. According to him, minimum energy configuration is obtained when nine wave functions ($(3d)^5, ~(4s)^1, ~(4p)^3$) are combined to produce nine hybrid wave functions ($spd$ hybridization). Out of nine, six have conductive hybrid orbital character (these form extended states from atom to atom running throughout the lattice), and remaining three have localized atomic character. Then he postulates Zener type mechanism. There is Hund's coupling between electrons in localized (atomic) hybrid orbitals and electrons in conductive hybrid orbitals. This Hund's coupling tends to align the spins of electrons in the conductive hybrid orbitals and of electrons in localized orbitals, thereby leading to spin polarization and ferromagnetism. He neglected the direct or inter-atomic exchange interaction between adjacent atoms (i.e., $J$ term in the VZ model is not there). Pauling's model have similar problems as that of VZ, and the division of d-orbitals as postulated by Pauling is never observed experimentally. But, according to Anderson\cite{8},  these valence bond ideas find their way in high-$T_c$ cuprate superconductors.

\section{Debate and its resolution}

After discussing these developments which started along the approach of localized model, let us return back to the itinerant picture and to the debate between these two extreme pictures. As mentioned before, in 1950s when it was not clear whether d-electrons in transition metals are localized or itinerant it was a real problem to decide whether Heisenberg model is more appropriate to understand ferromagnetism of some transition metals, or the itinerant Stoner model is more appropriate. Both approaches have their pros and cons. We discuss them one by one:

\subsection{In favor of itinerant (Stoner) model}

Stoner model is conceptually elegant and computationally easy to implement. It qualitatively reproduced magnetism versus temperature graph. The phenomenologically introduced exchange interaction by Stoner finds its justification in Slater's intra-atomic exchange and Hund's mechanism. However, this justification is open to criticism (refer to part A). The most important success of Stoner model is that it can address fractional Bohr magneton numbers found in saturation magnetization (refer to table 1). 

\subsection{Against itinerant (Stoner) model}

Ferromagnetic metals obey Curie-Weiss (CW) law (linear graph between inverse susceptibility and temperature) to a reasonably good approximation. The plot of inverse susceptibility versus temperature from Stoner model shows appreciable curvature, instead of being linear. Thus it fails to reproduce CW law.  Also the calculated values of $T_c$ for a reasonable value of exchange parameter extracted from the spectroscopic data is an order of magnitude higher than experimental value. Thus Stoner model also fails to reproduce the value of the Curie temperature. 

Another drawback of the Stoner model is that electronic correlation effects are completely neglected. As discussed in the next section full itineracy requires momentarily arbitrary ionization state of a given atom. This costs correlation energy. For example, free state of iron atom has the configuration $[Ar]3d^64s^2$. s-orbitals have extended wave functions and in the condensed state they overlap considerably to form wide s-bands. d-orbitals overlap comparably less, and form narrow d-bands. Fully itinerant d-electrons, according to Stoner model, can leave a variety of d-orbital configurations on a given atom: $d^8,~d^7,~d^6,~d^5,~d^4$ etc. In Stoner model they are all equally probable. But this of course wrong as states $d^8$ and $d^4$ will have higher energies due to Coulomb repulsion between electrons (two extremes of the ionization states). Therefore these configurations are very unlikely to occur, but Stoner model implicitly assumes that these occur with equal probability.  Roughly speaking Coulomb repulsion acting between opposite spin electrons lead to correlation energy. This is completely neglected in the Stoner model. And it is one of the major drawbacks of the model (refer to section VIII).

\begin{table}[h!]
\caption{Heisenberg versus Stoner}
\begin{center}
  \begin{tabular}{ |p{3cm} || p{3cm}| }
      \hline\hline
    Heisenberg Model  &  Stoner model \\  
    \hline
 Heisenberg  model treats localized electron cases and it is a generalization of the Heitler-London approach.  &  Stoner model is for itinerant electron problems and it is a generalization of the Bloch approach.  \\
    \hline
   It always give integral number of Bohr magnetons for the saturation intensity. & Stoner theory can explain saturation intensity with fractional Bohr magnetons. \\
    \hline
    It can address Curie-Weiss law for magnetic susceptibility. & It cannot explain the Curie-Weiss law adequately. And calculated values of $T_c$ are too high  \\    \hline\hline
  \end{tabular}
  \end{center}
\end{table}

\subsection{In favor of localized (Heisenberg) model}

The most important success of the Heisenberg model is that it can address CW law in an elegant way. Due to this fact before the resolution of the debate, it was the model of choice to analyze experimental data even in metals. Later on it became evident that this most suited for insulators, and became the seed to further developments in the field of magnetism of insulator compounds (next section).

\subsection{Against localized (Heisenberg) model}

The biggest drawback of the Heisenberg model is that it  always give integer number of Bohr magnetons for saturation magnetization. This result of the model is at variance with observed facts. Also, as discussed previously, the sign of the exchange integral $J$ is a vexed issue. The Slater curve provides a rule of thumb  but no theoretical derivation of it exists. 

From the above discussion it is clear that Heisenberg model is not appropriate at all to discuss ferromagnetism of metals. The Stoner model seems appropriate but it has serious drawbacks in that (1) it completely neglects correlation effects and (2) it is not able to capture thermodynamical properties at finite temperatures. Before we enter into further developments along itinerant (stoner) approach which rectify the above drawbacks, we would like to briefly brush-up the well settled issues of magnetism in insulator compounds in the next section. 

\section{Exchange interactions in insulator compounds}

Magnetism and exchange interactions in insulator compounds  are comparably well understood topics, and are well treated in the literature (refer for example to\cite{1,6,9}). Thus our discussion of this topic here is brief. The key point of the issue of magnetism in insulator compounds was clarified by Mott and Anderson in 1950s, and it can be stated in the following way. According to Bloch-Wilson theory of energy band formation in crystalline materials, materials with completely filled or completely empty bands are insulators while materials with partly filled bands are metals. It turns out, as first pointed out by Mott\cite{10}, that it is only the half-truth, not true in all the cases of insulating behavior. Insulating behavior can be due to a different mechanism and most importantly it leads to magnetic properties. Consider for concreteness the example of $NiO$. In this compound $Ni$ is in the valence state $Ni^{2+}$ with only valence 8 electrons partly filling the $d$ band.  The partly filled d band according to Bloch-Wilson picture should lead to conduction. However, as is now well known, there is strong electron correlation in the narrow d bands which energetically prohibits the flit (hop) of an electron from one $Ni$ atom to another, thus giving insulating behavior. Such insulators are often magnetic as localized electrons carry a magnetic moment. These magnetic-insulator compounds are now called Mott-Anderson or Mott insulators. Magnetism of such systems was elaborated, among other investigators, by Anderson\cite{9}. 

Magnetic behavior in insulating systems originates from a variety of exchange interactions. Below we discuss the main ones:
\begin{enumerate}
\item Direct exchange
\item Superexchange
\item Double exchange
\end{enumerate}
We briefly discuss each of them. Detailed expositions can be found in\cite{1,9}.

Direct exchange is basically the Heisenberg exchange between two nearby spins as discussed in section.  There we noticed that it is not the exchange mechanism in metals. In insulator compounds also it is not the most common one. One can cite the example of $MnF_2$ where direct exchange is thought to be operative. The exchange integral $J$ in this case is negative and leads to anti-ferromagnetism with $T_N\simeq10~K$. 

The most common exchange mechanism in insulating magnets is the superexchange mechanism first pointed out by Kramers and elaborated, among others, by Anderson.  To illustrate the basic principle of the mechanism consider the example of manganese oxide ($MnO$). In the simplest structural unit two $Mn$ cations are bounded together by one oxygen anion in the center.  Denote two metals ions by $M_1$ and $M_2$ and spin states by $M_1^{\uparrow}-O^{\uparrow\downarrow}-M_2^{\downarrow}$. There are two states $state_1 =M_1^{\uparrow}-O^{\uparrow\downarrow}-M_2^{\downarrow} $ and $state_2 = M_1^{\downarrow}-O^{\uparrow\downarrow}-M_2^{\downarrow}$, one with antiparallel spins on metal atoms and the other with parallel spins. If there is no coupling between the spin states on two metal cations, then these states will be degenerate (will have the same energy) and form the ground state. An excited state will involve a virtual transfer of an electron from central anion to its nearby cation, for example, transferring one electron from $O$ to $M_1$. Now according to the quantum mechanical theory of resonance phenomenon first elaborated by Pauling, the excited state wave functions combine with the ground state wave function to produce a hybrid state with even lower energy  thus stability\footnote{This is the most fundamental principle of the chemical bond theory\cite{10a}.} (lower than the above mentioned degenerate ground states). But this also removes the degeneracy. An excited state in the above example corresponds to transfer of an electron from ligand atom to its neighbor metal atom ($M_1^{\uparrow\uparrow}-O^{\downarrow}-M_2^{\downarrow}$). If the valence orbital on the metal atom is less than half full, then the transferred electron will align parallel to already existing spin on $M_1$ (i.e., the Hund's rule). Similar mechanism happen on $M_2$.  Thus it is clear that $state_1$  with antiparallel spins on metal atoms will have lower energy, and leads to antiferromagnetism. In the oxygen p orbitals electrons stay paired due to Pauli principle. Thus it is the Hund mechanism operating in metal atoms and virtual transfer of electrons from the ligand atom that leads to antiferromagnetism. A perturbational calculation was done by Anderson to calculate the magnitude of the effect, and there are other details. Our discussion here is restricted to be semi technical and brief, for more details readers are advised to refer to\cite{1,9}.

The double exchange mechanism is thought to be operative in mixed valency systems, and was first proposed by Zener\cite{4,11}. Consider a system in which two configurations ($M^+$, and $M^{++}$) of the metal ion are possible. For example in crystals of $La MnO_3$ both $Mn^{3+}$ and $Mn^{4+}$ co-exist.  Then the transitions between two states $M^{++}-O^{--}-M^{+}$ and $M^{+}-O^{--}-M^{++}$ can occur i.e., valency of the ligand atom remains the same, but valency of the metal atoms fluctuates. Now watch out the spin states on metal atoms $M_1^{\uparrow\uparrow}-O^{\uparrow\downarrow}-M_2^{\uparrow}$ with Hund's rule operating on metal atoms (orbitals less than half full and ferromagnetic spin alignment). Suppose that one up spin electron transfers from $O$ to $M_2$ and simultaneously an up spin electron transfers from $M_1$ to $O$ giving:  $M_1^{\uparrow}-O^{\uparrow\downarrow}-M_2^{\uparrow\uparrow}$ i.e., there is a double exchange (from $M_1$ to $O$ and from $O$ to $M_2$). Then by the general theory, resonance between states $M_1^{\uparrow\uparrow}-O^{\uparrow\downarrow}-M_2^{\uparrow}$ and  $M_1^{\uparrow}-O^{\uparrow\downarrow}-M_2^{\uparrow\uparrow}$ leads to lower energy and thus more stability. It can be easily checked that this mechanism will not operate if spins on metal atoms are anti-parallel.  In conclusion, this mechanism stabilizes the ferromagnetic spin arrangement\cite{1,4,11}.  

After this brief discussion of the exchange interactions in insulator compounds, we return to the question of the correlation effects in the itinerant (Stoner) model.

\section{Van Vleck again and his "middle-road" theory}

As discussed before there are two extreme models of ferromagnetism. One is localized Heisenberg model also called the non-polar model in which polarity of atoms stay constant in time. Electron migration from one atom to another is prohibited. The other extreme is the itinerant model where electron migration from one atom to another is the main feature.  This is also called the polar model, as polarity of an atom is ever changing with time. At one instant it may have more electrons than the average and at the other instant less electrons than the average number. 

To make the discussion concrete consider the example of metal nickel. Nickel atom in free state has the configuration $[Ar]3 d^8 4s^2$. In condensed state $4s$ wave functions of adjacent atoms overlap considerably and form a wide conduction $s$ band. $3d$ wave functions overlap comparably weakly and form a narrow $d$ band. If we assume, as van Vleck\cite{12} assumed, that 70 percent of the wide $4s$ band is above the $3d$ band, then the minimum energy configuration  will be obtained when some of the $4s$ electrons migrate to $3d$ band.  Most stable configuration corresponds to the case when on the average $2 \times 0.7 = 1.4$ electrons per atom migrate to $3d$ orbitals making their average occupancy to $8+1.4 = 9.4$. The average occupancy of a $4s$ orbital will be $2-1.4 = 0.6$. Thus the minimum energy configuration symbolically can be written as $[Ar] 3 d^{1.4} 4s^{0.6}$. This could actually imply that there are two types of configurations: $[Ar] 3D^94s^1$ on 60 percent of atoms, and $[Ar] 3d^{10} 4s^0$ on 40 percent of atoms, making the average configuration $[Ar] 3d^{9.4} 4s^{0.6}$. Define $A\equiv [Ar] 3D^94s^1$ and $B \equiv [Ar] 3d^{10} 4s^0$, and  consider two extreme situations: one is that the configurations A and B on a given set of atoms are permanent  in time. That is if an atom is in configuration A it will remain in it (no electron migration from the atom or into the atom takes palce). This "stagnant" situation corresponds to the Heisenberg non-polar model. On the other extreme end which corresponds to the itinerant or polar picture electron migration is freely allowed, and all the configurations like $d^{10},~d^{9},~d^{8},~d^7,~etc$ are possible with equal ease. This situation corresponds to the Stoner model. Clearly, ionization energies of the states  $d^{10},~d^{9},~d^{8},~d^7,~etc$ differ, and this fact is completely neglected in the Stoner theory as mentioned before. In other words these "correlation effects" are completely neglected in Stoner theory.

\begin{figure}[!h]
\begin{center}
\includegraphics[height=5cm]{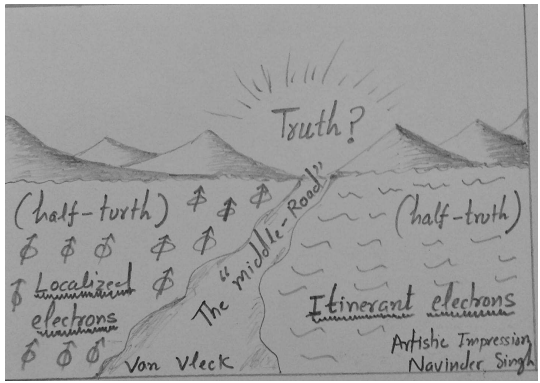}
\caption{An artistic impression of the "middle-road" theory.}
\end{center}
\end{figure}

In1940, van Vleck with his student Hurwitz, proposed a "middle-road" theory. As the name suggests Vleck-Hurwitz theory avoids both the extreme situations (localization and fully uncorrelated itineracy). In their view the configurations A or B on a given atom are not permanent in time, rather configuration of a given atom fluctuates between A and B (that is with minimum polarity). And on the whole at a given instant 60 percent of atoms will have configuration A and 40 percent will have configuration B. They performed rough estimates along these lines, but detailed calculations were not feasible due to the complexity of the problem\cite{12}, and the success was partial.  In conclusion, these were the "first steps" of incorporating correlation effects in the Stoner theory. We would like to end our presentation of the "middle-road" theory with the appropriate words of Kubo and Nagamiya\cite{6a}:

".....Many investigators agree in that this is most desirable. But it is still very difficult to pave this road......"

\section{The Friedel-Alexander-Anderson-Moriya theory of moment formation in pure iron group metals}

Building on the above ideas of van Vleck on correlation effects in transition metals, Anderson developed the theory of moment formation in a magnetic impurity atom in a metal (as dilute alloy of $Mn$ in $Cu$). There is an extensive literature on this very important and interesting topic of Anderson impurity problem\cite{6,13,13a,14,15} and we will not go into this here, rather we will continue our discussion of the traditional and complicated problem of ferromagnetism in iron group metals (pure metal not alloys) within the itinerant picture but including electron correlation effects.

The original ideas go back to van Vleck (the middle-road theory) and to Friedel (virtual bound states formation in dilute magnetic alloys). Anderson argued that the idea of virtual bound states can be applied to local moment formation in pure ferromagnetic metals. Thus an effective Heisenberg picture becomes emergent and the problems of the itinerant model in explaining the Curie-Weiss law can be resolved using an effective Heisenberg model for metals! It appear counter-intuitive to imagine local moments emerging from itinerant electrons. There is a very physical picture, due to Cyrot\cite{16}, to understand how local moments can form in an otherwise itinerant model. In Cyrot's words:

"The spin of an atom, i.e., the total spin of all the electrons on that atom, fluctuates randomly in magnitude and
direction. What effect might one expect from the electron interaction? We recall that Hund's first rule for atoms indicates that the intra-atomic interactions will aline the electron spins on an atom. We might expect a tendency to produce the same result in a metal, since if an atom has a spin up it will tend to attract electrons with spins up and repel those with spins down. On this account one would suppose that the total spin on an atom at any one instant tends to be self-perpetuating, so that the spin value can persist for a period long compared to the d-electron hopping time. The electrons on the atoms are always changing about, due to the band motion, but the magnetic moment of the atom persists due to the correlated nature of the electrons' motion. In these circumstances, one can consider the spin as being associated with the atom rather than with the individual electrons. Here we see the possibility of an atomic or a Heisenberg model emerging from the effect of correlations in the band model."   

Cyrot's words are sufficiently clear and we cannot add to it more. In conclusion, electron correlation leads to quasi-localization. Next comes the question of exchange interactions in these "induced" moments. This was addressed by Alexander and Anderson\cite{17} and by Toru Moriya\cite{18,19}. These investigations showed that instead of the Heisenberg's $J$ which is given by the Slater curve, one obtains an effective exchange interaction between the induced moments which follows a different rule. The effective exchange interaction is of ferromagnetic nature when atomic d shell is either nearly empty or nearly full, otherwise the interaction is antiferromagnetic. 

This line of approach was further extended by the introduction of powerful functional integral methods by Schriefer and by Cyrot (refer to \cite{19a}). Although this approach via strong correlation seemed quite promising but it fails to address weakly ferromagnetic systems like $ZrZn_2,~Sc_3In$ etc. Thus it could not provide a comprehensive picture. It turns out that a completely new mechanism for CW law operates in these materials (or may be in iron group metals) in which moments are not localized in real space rather they are localized in momentum space.  Such an approach was advanced by Moriya and kawabata in 1970s, and it is know known as the Self-Consistent-Renormalization (SCR) theory. This theory takes into account the correlation effects beyond mean field theory and in addition takes into account the renormalization effects  of spin fluctuations on the equilibrium state. Our next article in the series is devoted to the SCR theory and other theories that take correlation effects into account. For a detailed account of the SCR theory consult\cite{19a}. To consider the effect of spin fluctuations on thermodynamical properties, another theory along the lines of Landau theory of phase transitions was developed by Lonzarich and Taillefer\cite{20}.

\section{Conclusion}

The following lines by Toru Moriya are sufficiently clear to sort out the debate between localized and itinerant pictures:

".... the magnetic insulator compounds and rare earth magnets are described in terms of the localized electron model, while the ferromagnetic d-electron metals should be described on the itinerant electron model {\it with the approximation method beyond the mean field level, properly taking account of the effects of electron-electron correlation....} "

-----Toru Moriya\cite{21}.

When it became clear (from 1960s onwards) that d electrons in iron group metals are to be treated as itinerant electrons, the main aim of the ensuing investigations was to incorporate electron correlation effects into the itinerant (Stoner) picture, and to resolve its problems. Several investigators contributed in this important development. The investigations of Kanamori, Gutzwiller, Hubbard, Cyrot, Moriya, Kawabata, and Okabe (among others) are important ones, and will be reviewed in next paper  this series. 

In this paper, the difficult topics of exchange and correlation in itinerant and localized models are explained using semi-technical language. In summary, early attempts were focused on obtaining ferromagnetic exchange in the Heisenberg model. The Slater curve provided a provisional picture or a rule of thumb. The Vonsovsky-Zener model suggested a mechanism (s-d exchange) which could in principle provide the required ferromagnetic exchange, however it was found later on that it could not be applied to the iron group metals (due to itinerant nature of the d electrons) and was discarded. But it lead to other very interesting developments in f-electron and magnetic impurity systems (RKKY interaction and the Kondo effect). Pauling tried to develop a theory of ferromagnetism in iron group metals using his valence bond ideas, but this was also not successful as discussed in the text. The debate between the itinerant model and localized model was resolved and the main conclusion from 1960s onwards was to properly incorporate the correlation effects in the itinerant model. Early attempts were made by van Vleck in this direction with his "middle-road" theory or minimum polarity model. But to build a quantitative theory along these lines turned out to be hard. The Friedel-Alexander-Anderson-Moriya theory of moment formation in pure iron group metals builds upon these ideas.  As discussed, the problems of these approaches set the stage for the arrival of the SCR theory. In author's opinion one of the most successful development is the introduction of the SCR theory by Moriya and kawabata which goes beyond the Hartree-Fock and random phase approximations in treating the correlation effects, and in addition, it takes into account the effect of thermal excitations of magnetic nature (i.e., thermal spin fluctuations) on the equilibrium state i.e., the renormalization aspect of the SCR theory. These theories form the subject matter of our next article in this series.
%----------------------------------------------------------------------------------------
%	REFERENCE LIST
%----------------------------------------------------------------------------------------

%----------------------------------------------------------------------------------------

\end{document}